\documentclass{article}

\usepackage{PRIMEarxiv}
\usepackage[utf8]{inputenc} 
\usepackage[T1]{fontenc}    
\usepackage{hyperref}       
\hypersetup{pdfborder = {0 0 0}}
\usepackage{url}            
\usepackage{booktabs}       
\usepackage{amsfonts}       
\usepackage{nicefrac}       
\usepackage{microtype}      
\usepackage{lipsum}
\usepackage{fancyhdr}       
\usepackage{siunitx}

\usepackage{graphicx}       
\usepackage{float}
\usepackage[export]{adjustbox}

\usepackage{cite}

\usepackage{todonotes}
\usepackage{amssymb, amsmath}
\graphicspath{{media/}}     

\usepackage{titlesec}
\titlespacing*{\section}{\baselineskip}{\baselineskip}{0.5\baselineskip}
\titlespacing*{\subsection}{\baselineskip}{\baselineskip}{0.5\baselineskip}

\usepackage{xspace}
\newcommand{\PYTHIA} {{\textsc{pythia}}\xspace}
\newcommand{\DELPHES} {{\textsc{delphes}}\xspace}

\newcommand{\MeV}{\ensuremath{\,\text{Me\hspace{-.08em}V}}\xspace}
\newcommand{\GeV}{\ensuremath{\,\text{Ge\hspace{-.08em}V}}\xspace}
\newcommand{\TeV}{\ensuremath{\,\text{Te\hspace{-.08em}V}}\xspace}

\newcommand{\et}{\ensuremath{E_{\mathrm{T}}}\xspace}
\newcommand{\HT}{\ensuremath{H_{\mathrm{T}}}\xspace}

\newcommand{\phid}{\ensuremath{\phi_{\mathrm{D}}}\xspace}
\newcommand{\zd}{\ensuremath{Z_{\mathrm{D}}}\xspace}

\title{Autoencoders for Real-Time SUEP Detection}
\author{Simranjit Singh Chhibra$^{2,4}$,
        Nadezda Chernyavskaya$^{2}$, 
        Benedikt Maier$^{2,3}$, 
        Maurzio Pierini$^{2}$,
        Syed Hasan$^{1,5}$
        \\        
        $^{1}$ETH Zurich, Switzerland\\
        $^{2}$European Organization for Nuclear Research (CERN), Switzerland\\
        $^{3}$Karlsruhe Institute of Technology (KIT), Germany\\
        $^{4}$Queen Mary University of London (QMUL), UK \\
        $^{5}$Scuola Normale Superiore (SNS) di Pisa, Italy
    }

\begin{document}

\maketitle

\begin{center}
    \href{to:benedikt.maier@cern.ch}{benedikt.maier@cern.ch}
\end{center}

\vspace{2cm}

\begin{abstract} 
Confining dark sectors with pseudo-conformal dynamics can produce Soft Unclustered Energy Patterns (SUEP), at the Large Hadron Collider: the production of dark quarks in proton-proton collisions leading to a dark shower and the high-multiplicity production of dark hadrons. The final experimental signature is spherically-symmetric energy deposits by an anomalously large number of soft Standard Model particles with a transverse energy of $\mathcal{O}(10^2)\MeV$. Assuming Yukawa-like couplings of the scalar portal state, the dominant production mode is gluon fusion, and the dominant background comes from multi-jet QCD events. We have developed a deep learning-based Anomaly Detection technique to reject QCD jets and identify any anomalous signature, including SUEP, in real-time in the High-Level Trigger system of the Compact Muon Solenoid experiment at the Large Hadron Collider. A deep convolutional neural autoencoder network has been trained using QCD events by taking transverse energy deposits in the inner tracker, electromagnetic calorimeter, and hadron calorimeter sub-detectors as 3-channel image data. Due to the sparse nature of the data, only $\sim$0.5\% of the total $\sim$$\SI{300}{k}$ image pixels have non-zero values. To tackle this challenge, a non-standard loss function, the inverse of the so-called Dice Loss, is exploited. The trained autoencoder with learned spatial features of QCD jets can detect 40\% of the SUEP events, with a QCD event mistagging rate as low as 2\%. The model inference time has been measured using the {\texttt{Intel\textregistered~Core$^{\text{TM}}$~i5-9600KF}} processor and found to be $\sim$$\SI{20}{\milli\second}$, which perfectly satisfies the High-Level Trigger system's latency of $\mathcal{O}(10^2)~\SI{}{\milli\second}$. Given the virtue of the unsupervised learning of the autoencoders, the trained model can be applied to any new physics model that predicts an experimental signature anomalous to QCD jets.
\end{abstract}

\newpage

\section{Introduction}
Hidden valleys, referring to hidden sectors with a confining gauge group, can produce dark quarks in proton-proton collisions at the Large Hadron Collider (LHC), leading to a dark shower and the production of a large number of dark hadrons (\phid), analogous to QCD jets~\cite{Strassler:2006im,Knapen:2016hky,Barron:2021btf}. Depending on the details of the theory, the dark showers can follow large-angle emission, and dark hadrons do not arrange in narrow QCD-like jets. The decay of dark hadrons results in dark photons (\zd), which further decay to low-energy Standard Model particles with transverse energy (\et) of $\mathcal{O}(10^2)$\MeV, with the final experimental signature being high-multiplicity spherically-symmetric Soft Unclustered Energy Patterns, or SUEPs. We focus on a well-motivated scenario where SUEP is produced in exotic Higgs ($H$) boson decays via gluon-gluon fusion and all dark hadrons decay promptly and exclusively to pions and leptons---an experimental nightmare scenario because of an overwhelming multi-jet QCD background. 

We have developed an Anomaly Detection (AD) technique~\cite{darkmachines, lhco2020} exploiting unsupervised deep learning for real-time SUEP detection in the High-Level Trigger (HLT) system~\cite{CMS:2016ngn} of the Compact Muon Solenoid (CMS) experiment~\cite{Chatrchyan:2008zzk} at the LHC\footnote{The CMS trigger system is a two-tiered system. The first level (L1) triggers, implemented on custom electronics, process information from the calorimeters and muon system and select $\SI{100}{\kilo\hertz}$ events of physics interest from $\SI{40}{\mega\hertz}$ collision events within a time interval of $\SI{4}{\micro\second}$. The second level (HLT) triggers run an optimised version of full-event reconstruction software on a farm of processors and further reduce the event rate to $\SI{1}{\kilo\hertz}$ with a latency of $\mathcal{O}(10^2)~\SI{}{\milli\second}$.}. Most AD techniques developed for High-Energy Physics applications rely on the online or offline reconstruction of collision events, which could be ineffective for reconstructing new physics signatures. We utilise raw quantities reconstructed in the HLT system, i.e., energy deposits in different sub-detectors (input for the online particle reconstruction). We have trained a deep convolutional neural autoencoder network (ConvAE) using QCD events, which is designed to detect events significantly different from the training data without prior knowledge of the signal characteristics. The data are three-channel images: two-dimensional (in the $\eta$-$\phi$ plane, where $\eta$ is the pseudorapidity) \et deposits in the inner tracker, electromagnetic calorimeter (ECAL), and hadron calorimeter (HCAL) sub-detectors of CMS. Even though our work targets SUEP event tagging, the trained model could identify any other non-QCD-like signature, such as semi-visible jets~\cite{Kar:2020bws}  and emerging jets~\cite{Schwaller:2015gea}, spanning a large part of the landscape of the hidden valley. We plan on covering those new physics signals in our next paper.

This paper is structured as follows: Section~\ref{Datasets} contains all the information about the QCD and SUEP event generation, simulation, and reconstruction. The event preselection is highlighted in the same section. The details about the ConvAE architecture and training are provided in Section~\ref{Autoencoders}. The trained ConvAE is applied to the QCD background and SUEP signal considering several Higgs boson mediator mass scenarios, and the performance results are given in Section~\ref{Results}. Section~\ref{Conclusions} concludes the paper.

\section{Datasets}
\label{Datasets}
\subsection{Generation, simulation, and reconstruction} 
\label{Generation, simulation and reconstruction} 
Both multi-jet QCD and SUEP events are generated using \PYTHIA8.244~\cite{Sjostrand:2014zea}. Detector effects are simulated and events are reconstructed with \DELPHES 3.5~\cite{deFavereau:2013fsa} using the CMS detector configuration. Proton-proton collisions at the LHC Run-3 centre-of-mass energy of 13.6\TeV are considered, with approximately 50 collisions per bunch crossing (pileup). A pileup reduction is applied to the reconstructed tracks: the \et deposits in the inner tracker by the pileup tracks are removed. No pileup reduction is applied for calorimeters, as per the HLT system's configuration during Run 3~\cite{CMS:2016ngn}.

The model parameters for the QCD event generation are identical to those in~\cite{CMS:2021dzg}. For SUEP event generation, we used the {\texttt{SUEP\_Generator}} plugin~\cite{Knapen:2021eip}, and the following parameter settings have been used:  

\begin{itemize}
\item mediator masses ($m_{H}$) = 125\GeV, 400\GeV, 700\GeV, and 1000\GeV
\item dark meson mass ($m_{\phid}$) = 2\GeV
\item branching ratio of $\phid\to2 \zd$ = 100\%
\item dark photon mass ($m_{\zd}$) = 0.7\GeV
\item branching ratios of $\zd\to \pi^+\pi^-,$ $e^+e^-,$ $\mu^+\mu^-$ = 70\%, 15\%, 15\%
\item dark temperature ($T_{\mathrm{D}}$) = 2\GeV~\cite{Knapen:2016hky}
\end{itemize}

Please note that we denote SUEP with a mediator mass of $m_{H}$\GeV as SUEP($m_{H}$~\GeV) in the rest of the paper.

\subsection{Event preselection} 
\label{Event preselection}
A preselection criterion is applied to both QCD and SUEP events based on the detector geometry and the typical characteristics of QCD events. The $|\eta|$ of \et deposits is required to be less than 2.5, as defined by the inner tracker coverage. The scalar sum of \et of reconstructed electrons, muons, photons, and jets (denoted as \HT) in an event is required to be $>$500\GeV, which is a typical \HT requirement at the HLT after L1 trigger selection during Run 3. The event preselection efficiency is found to be $\sim$85\% for QCD and $\sim$2.1\%, $\sim$10.4\%, $\sim$18.4\%, and $\sim$22.7\% for SUEP(125\GeV), SUEP(400\GeV), SUEP(700\GeV), and SUEP(1000\GeV), respectively.

\begin{figure}[h]
\centering
\adjincludegraphics[clip, trim={{.21\width} 0 {.074\width} 0}, width=0.32\textwidth]{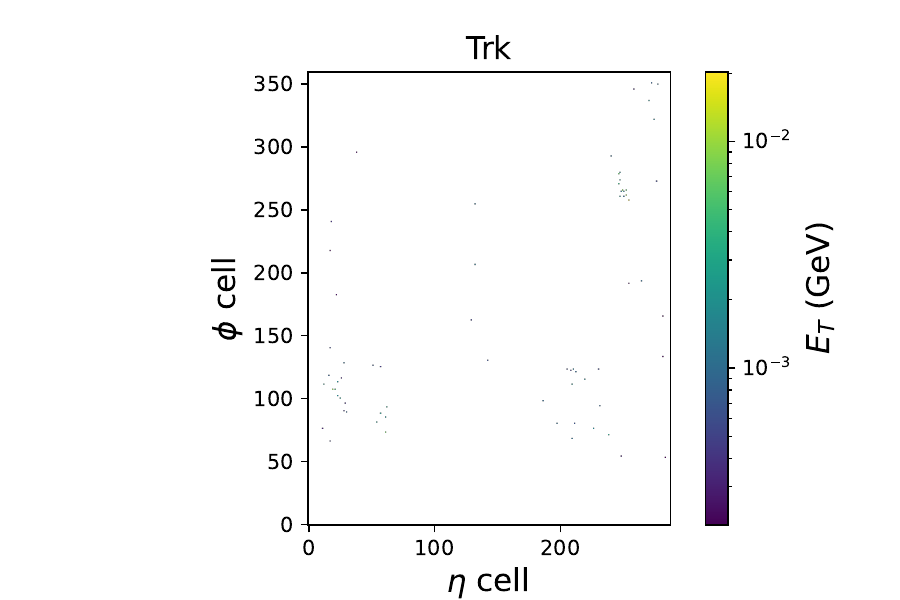}
\adjincludegraphics[clip, trim={{.21\width} 0 {.074\width} 0}, width=0.32\textwidth]{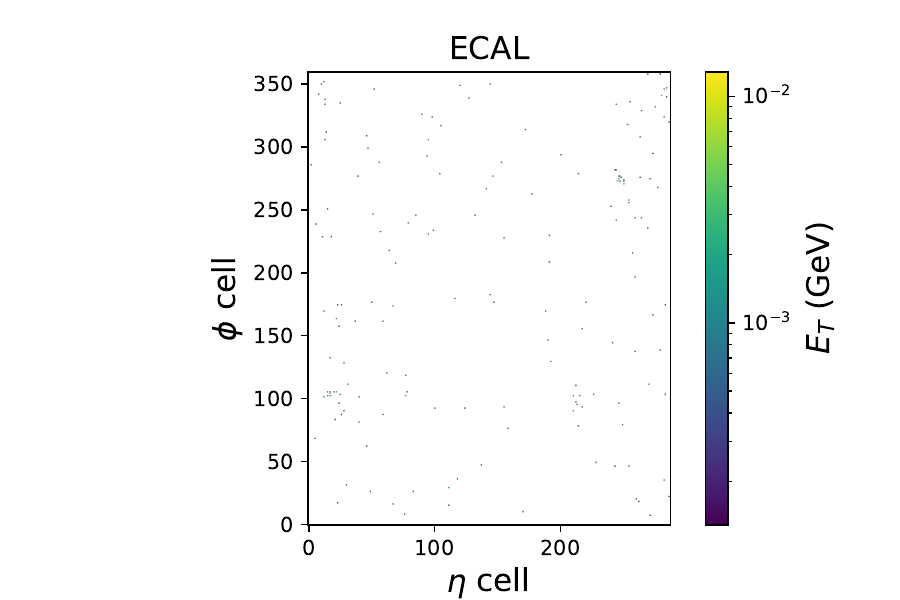}
\adjincludegraphics[clip, trim={{.21
\width} 0 {.074\width} 0}, width=0.32\textwidth]{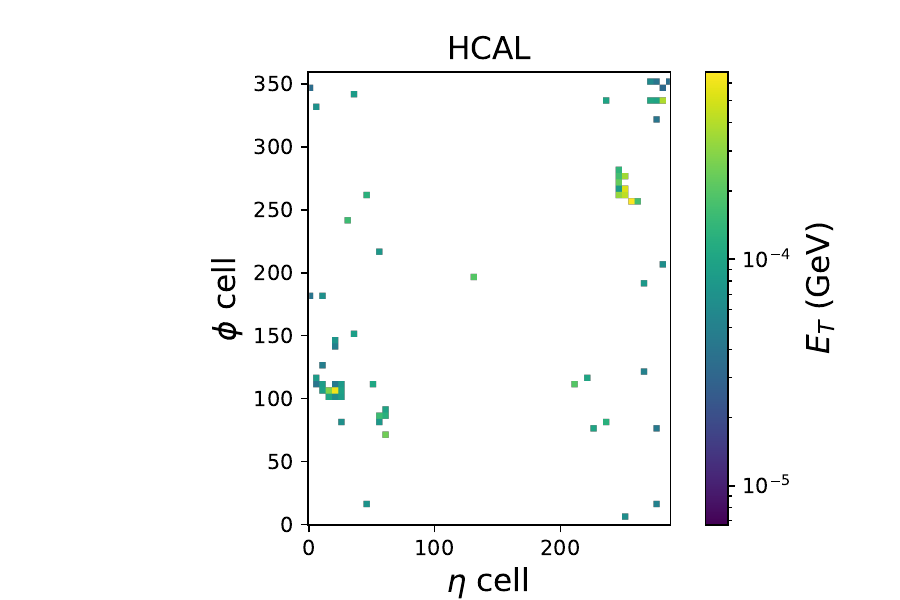}
\caption{\et deposits for a QCD event in the inner tracker (left), ECAL (middle), and HCAL (right).}
\label{fig:QCDInputEvent}
\end{figure}

\begin{figure}[h]
\centering
\adjincludegraphics[clip, trim={{.21\width} 0 {.074\width} 0}, width=0.32\textwidth]{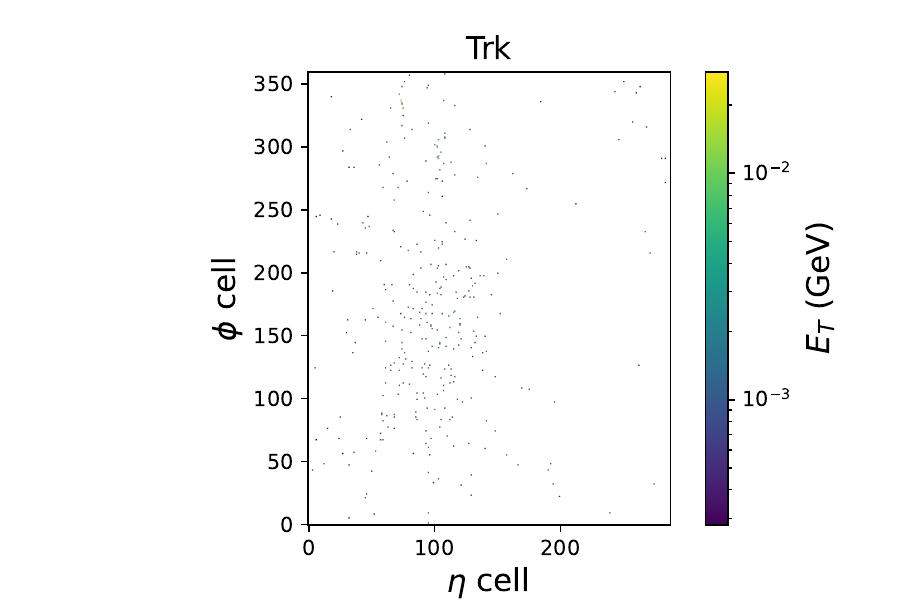}
\adjincludegraphics[clip, trim={{.21\width} 0 {.074\width} 0}, width=0.32\textwidth]{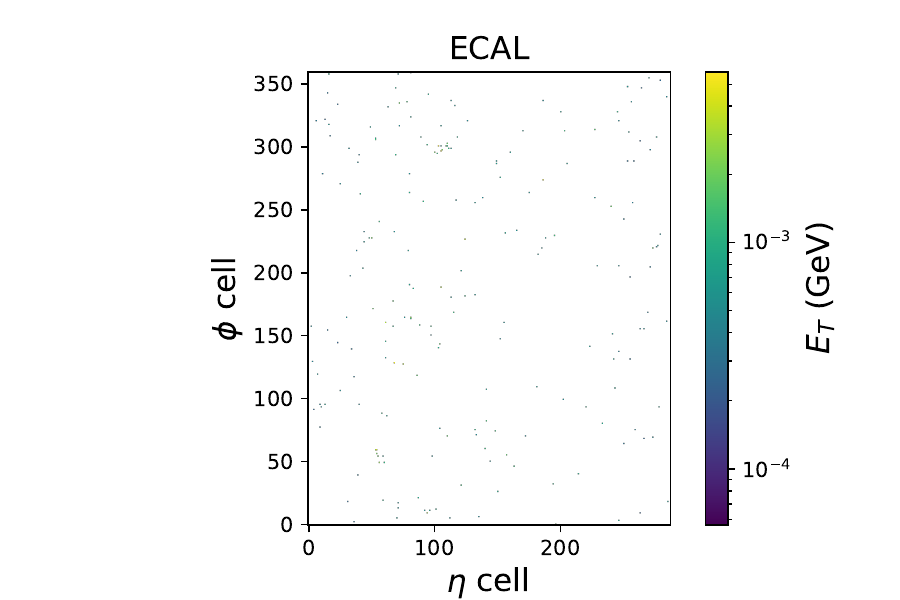}
\adjincludegraphics[clip, trim={{.21\width} 0 {.074\width} 0}, width=0.32\textwidth]{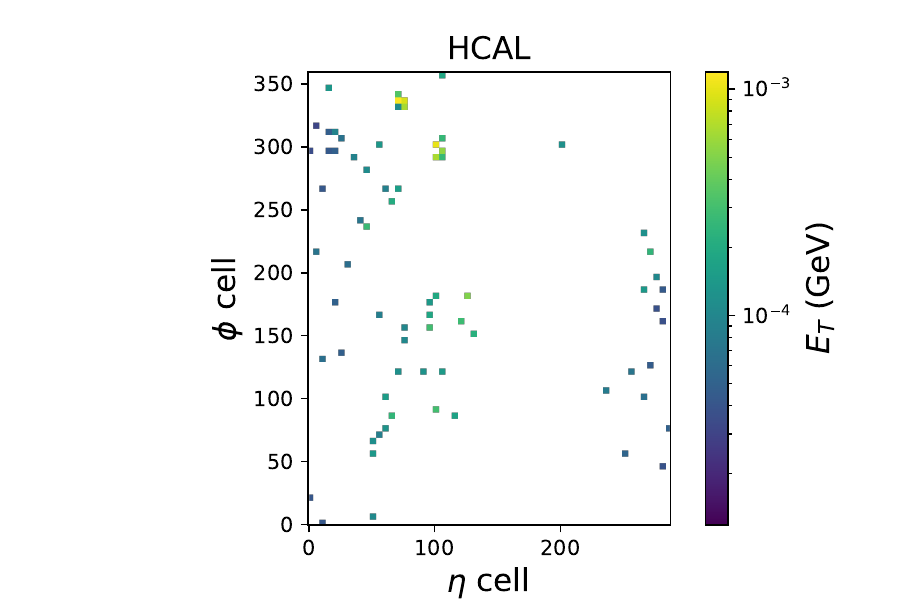}
\caption{\et deposits for a SUEP(1000\GeV) event in the inner tracker (left), ECAL (middle), and HCAL (right).}
\label{fig:SUEPInputEvent}
\end{figure}

The plots showing the \et deposits for a QCD event and a SUEP(1000\GeV) event, which satisfy the preselection criterion, in the inner tracker, ECAL, and HCAL are given in Figures~\ref{fig:QCDInputEvent} and~\ref{fig:SUEPInputEvent}, respectively. The ECAL is 25 times more granular (0.0174$\times$0.0174~$\SI{}{\radian^2}$ in $\eta$-$\phi$) than the HCAL (0.087$\times$0.087~$\SI{}{\radian^2}$ in $\eta$-$\phi$). Hence, each HCAL pixel is divided into 25 equal pixels to match the ECAL granularity. The final shape of the images is (288, 360, 3). For this particular QCD event, we can appreciate the typical signature of two jets in the $\eta$-$\phi$ plane. Also, we can acknowledge the sparse nature of the data---only $\sim$$\SI{15}{k}$ pixels out of a total of 3,11,040 image pixels have non-zero values. This observation is valid in general for all QCD events. On the other hand, for the SUEP(1000\GeV) event, there is a clear spherically-symmetric signature of a large number of low-\et particles in $\phi$. We normalised the \et values such that they are of $\mathcal{O}(1)$, which facilitates more stable and faster ConvAE training.

\section{Autoencoders}
\label{Autoencoders}
Our ultimate target is to reject QCD jets and identify anomalous signatures. The unsupervised learning nature of autoencoders (AEs) perfectly serves the purpose~\cite{Goodfellow-et-al-2016}. In its simplest form, an AE is a neural network (NN) that maps the input (ground truth) to a latent compressed representation (encoding), the so-called bottleneck, and then back to the input dimensionality (decoding), providing the output (reconstruction) as an approximation of the input. The quality of the AE training is measured in terms of a distance metric, the so-called reconstruction loss; e.g., for images, the loss function can be defined as the pixel-wise summed mean-squared difference between input and output. Once the AE is trained using background events with the objective of minimising the reconstruction loss, the model is expected to poorly reconstruct anomalous signatures and result in large reconstruction loss values. A threshold on the reconstruction loss can be imposed to tag anomalies.

\subsection{Architecture and training}
\label{Architecture and training}
The ConvAE architecture is detailed in Table~\ref{tab:AEArchitecture}. For encoding, there are five two-dimensional Convolutional NN (CNN)~\cite{journals/corr/OSheaN15} layers with a \texttt{kernel\_size} of (3, 3) and \texttt{strides} of (3, 3), (2, 2), or (2, 3). Up to the bottleneck, the input shape of (288, 360, 3) compresses to (6, 5, 8). The decoder part, which is exactly the mirror image of the encoder part, is composed of five two-dimensional Transposed CNN~\cite{dumoulin2016guide} layers with the same \texttt{kernel\_size} and exactly opposite \texttt{strides}. The {\texttt{padding}} is set to {\texttt{"same"}}. After each (Transposed) CNN layer, a \texttt{BatchNormalization} layer is introduced, which facilitates a more stable and faster training by bringing the mean of the output close to zero and the output standard deviation close to unity. The Parametric Rectified Linear Unit (\texttt{PReLU}) and Rectified Linear Unit (\texttt{ReLU})~\cite{he2015delving} activation functions are used for all the hidden layers and the output layer, respectively. This model architecture results in a total number of trainable hyperparameters equal to 3,574,065. The model was implemented in Keras/TensorFlow~\cite{chollet2015keras, tensorflow2015-whitepaper}, and it was trained for 100 epochs (with early stopping enabled) with a \texttt{batch\_size} of 128. The learning rate was set to $10^{-2}$, and the optimizer found to yield the best performance was \texttt{Adam}~\cite{kingma2014method}. The training was performed using the \texttt{NVIDIA\textregistered~Tesla\textregistered~V100~PCIe~32~GB} Graphics Processing Unit~\cite{Teslav100gpu}, and $\sim$25 minutes per epoch were consumed while model training.

\begin{table}[h]
\begin{center}
\caption{The ConvAE architecture.}
\label{tab:AEArchitecture}
\begin{tabular}{|c|c|c|c|} 
\hline
Layer(\texttt{kernel\_size}, \texttt{strides}) & Shape & \texttt{BatchNormalization} & Activation function \\ 
 \hline 
 \hline 
Input & (288, 360, 3) & & \\ 
 \hline 
 \hline 
\texttt{Conv2D((3, 3), (3, 3))} & (96, 120, 128) & yes & \texttt{PReLU} \\ 
 \hline 
\texttt{Conv2D((3, 3), (2, 2))} & (48, 60, 64)   & yes & \texttt{PReLU} \\ 
 \hline 
\texttt{Conv2D((3, 3), (2, 2))} & (24, 30, 32)     & yes & \texttt{PReLU} \\ 
 \hline 
\texttt{Conv2D((3, 3), (2, 2))} & (12, 15, 16)     & yes & \texttt{PReLU} \\  
 \hline 
\hline
\texttt{Conv2D((3, 3), (2, 3))} & (6, 5, 8)          & yes & \texttt{PReLU} \\
 \hline 
 \hline 
\texttt{Conv2DTranspose((3, 3), (2, 3))} & (12, 15, 16)     & yes & \texttt{PReLU} \\ 
 \hline 
\texttt{Conv2DTranspose((3, 3), (2, 2))} & (24, 30, 32)     & yes & \texttt{PReLU} \\ 
 \hline 
\texttt{Conv2DTranspose((3, 3), (2, 2))} & (48, 60, 64)   & yes & \texttt{PReLU} \\ 
 \hline 
\texttt{Conv2DTranspose((3, 3), (2, 2))} & (96, 120, 128) & yes & \texttt{PReLU} \\ 
 \hline 
\texttt{Conv2DTranspose((3, 3), (3, 3))} & (288, 360, 3)   & yes & \texttt{ReLU} \\ 
 \hline 
\end{tabular}
\end{center}
\end{table}

The biggest challenge we faced in accomplishing the task was data sparsity, as mentioned in Section~\ref{Event preselection}. The model had a huge tendency to learn the zeros rather than the non-zero values. The standard loss functions, such as Mean Squared Error~\cite{SammutWebb2017} and Cross-Entropy~\cite{crossentropy1}, did not work well. We exploited the inverse of the so-called Dice Loss~\cite{10.1007/978-3-030-01231-1_35} as the loss function, as given by equation~\ref{eq:invdiceloss}. The $x_i$ and $x_i'$ denote the $i^{th}$-pixel values for the input and output, respectively. The $N$ is the total number of pixels. The denominator term contains the pixel-wise product of input and output, which forces the output to have the same non-zero pixels as the input to minimise the reconstruction loss. On the other hand, the numerator term and minus 1 ensure that the loss function gets its minimum at zero if the output becomes equal to the input. 

\begin{equation}
\label{eq:invdiceloss}
L(x, x') = \frac{\sum_{i}^{N}{x_i^2} + \sum_{i}^{N}{x_i'^2}}{2\sum_{i}^{N}{x_i x_i'}} - 1
\end{equation}

We used two independent sets of $\SI{50}{k}$ QCD events for the model training and validation. During the model training, no overfitting was observed, and the loss minimisation was smooth and saturated after 57 epochs.

\section{Results}
\label{Results}
\subsection{Autoencoder reconstruction}
\label{Autoencoder reconstruction}
\begin{figure}[h]
\centering
\includegraphics[width=.4\textwidth]{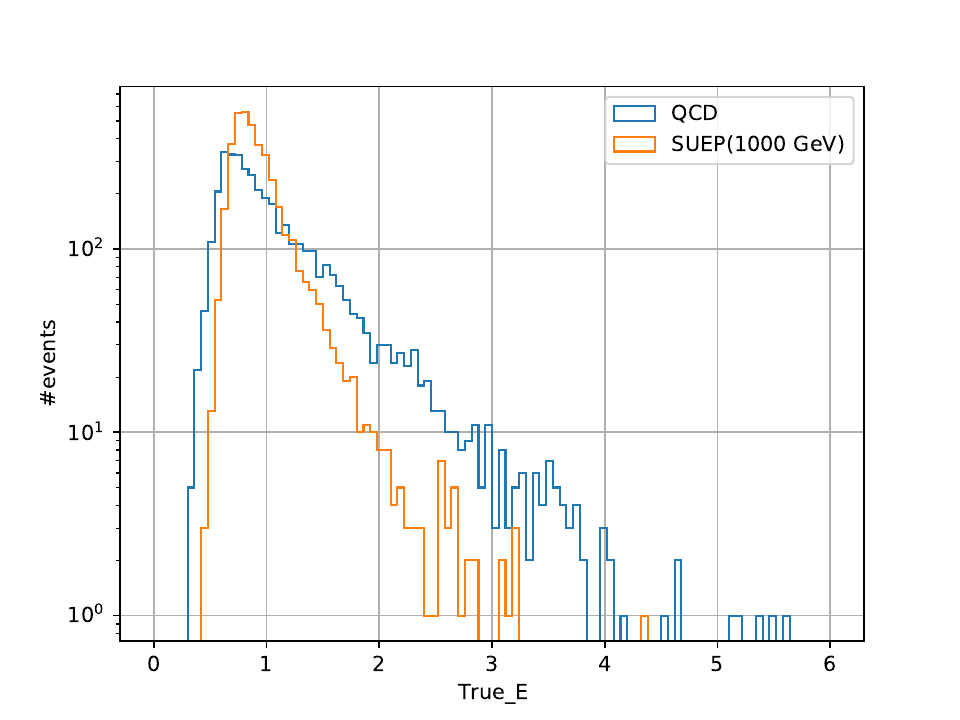}
\includegraphics[width=.4\textwidth]{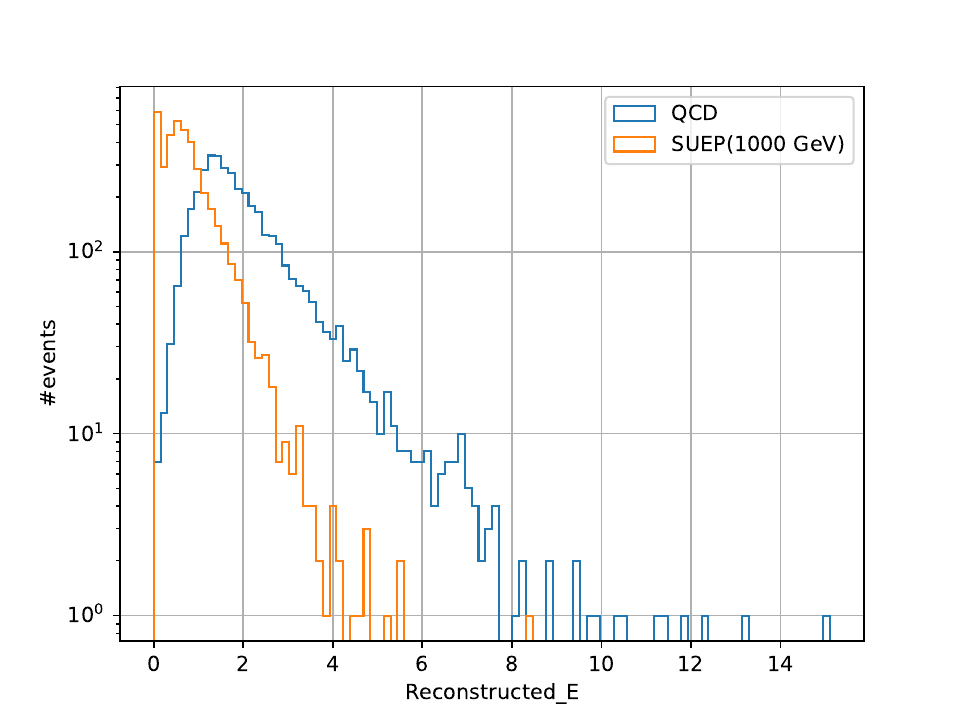}
\caption{Comparisons of true \et (left) and reconstructed \et (right) (sum over all the pixels) between QCD and SUEP(1000~\GeV).}
\label{fig:TrueAndRecoE}
\end{figure}

The trained ConvAE is tested using $\SI{4}{k}$ QCD and SUEP events each, which satisfy the event preselection criterion detailed in Section~\ref{Event preselection}. 
Figure~\ref{fig:TrueAndRecoE} represents a comparison of true \et (left) and reconstructed \et (right)---sum over all the pixels---between QCD and SUEP(1000~\GeV). 
It can be noted that the model reconstructs the \et for QCD events quite well but struggles for SUEP(1000~\GeV) \et reconstruction.

\begin{figure}[h]
\centering
\adjincludegraphics[clip, trim={{.21\width} 0 {.11\width} 0}, width=0.42\textwidth]{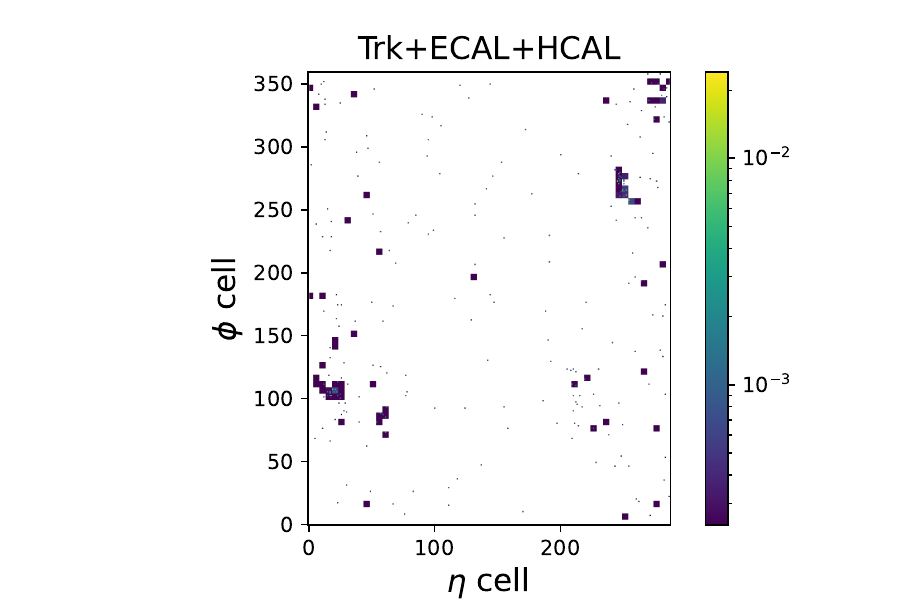}
\adjincludegraphics[clip, trim={{.21\width} 0 {.11\width} 0}, width=0.42\textwidth]{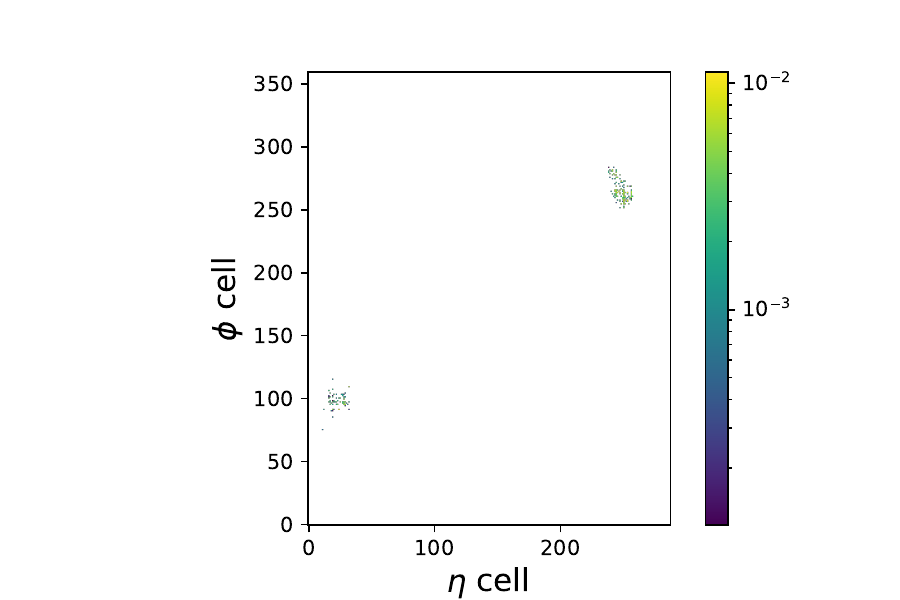}
\caption{Combined true \et (left) and reconstructed \et (right) for a QCD event in the inner tracker, ECAL, and HCAL.}
\label{fig:QCDReconstructedEvent}
\end{figure}

\begin{figure}[h]
\centering
\adjincludegraphics[clip, trim={{.21\width} 0 {.11\width} 0}, width=0.42\textwidth]{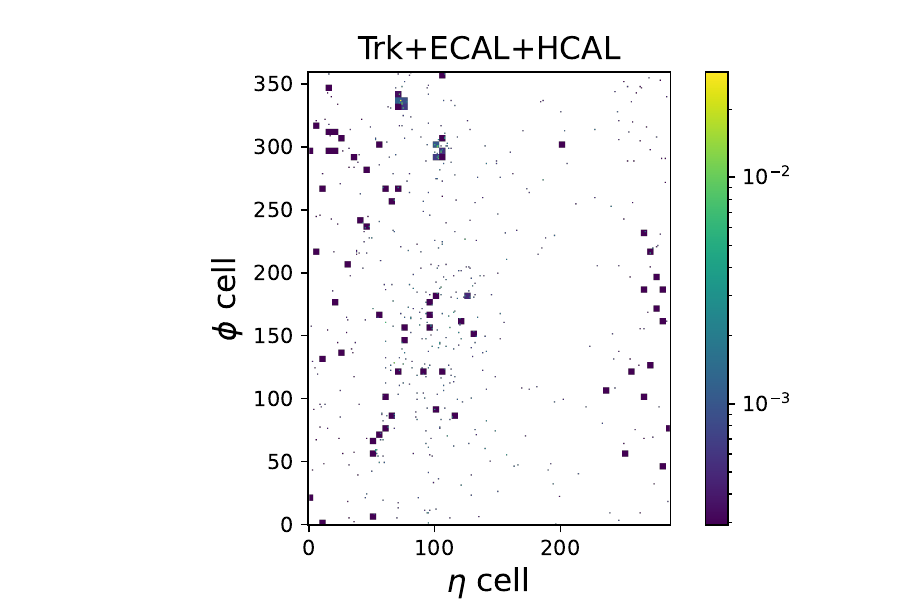}
\adjincludegraphics[clip, trim={{.21\width} 0 {.11\width} 0}, width=0.42\textwidth]{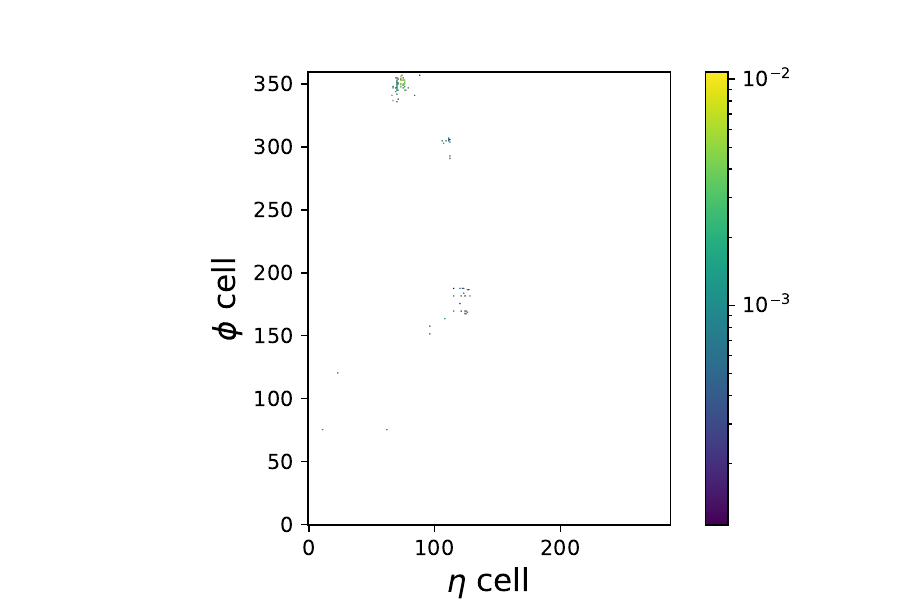}
\caption{Combined true \et (left) and reconstructed \et (right) for a SUEP(1000~\GeV) event in the inner tracker, ECAL, and HCAL.}
\label{fig:SUEPReconstructedEvent}
\end{figure}

The combined true \et for a QCD event in the inner tracker, ECAL, and HCAL is given in the left plot of Figure~\ref{fig:QCDReconstructedEvent} (same as Figure~\ref{fig:QCDInputEvent} but combined now). The reconstructed \et for the same event is shown in the right plot of Figure~\ref{fig:QCDReconstructedEvent}. The model is able to reconstruct the non-zero input pixels quite well. Similar plots for a SUEP(1000~\GeV) event are given in Figure~\ref{fig:SUEPReconstructedEvent}. In this case, the model reconstructed only the large \et clusters (in the top-left quadrant) but failed to reconstruct the whole spherically-symmetric signature of a large number of particles in $\phi$.

\subsection{Autoencoder performance}
\label{Autoencoder performance}
\begin{figure}[h]
\centering
\includegraphics[width=.4\textwidth]{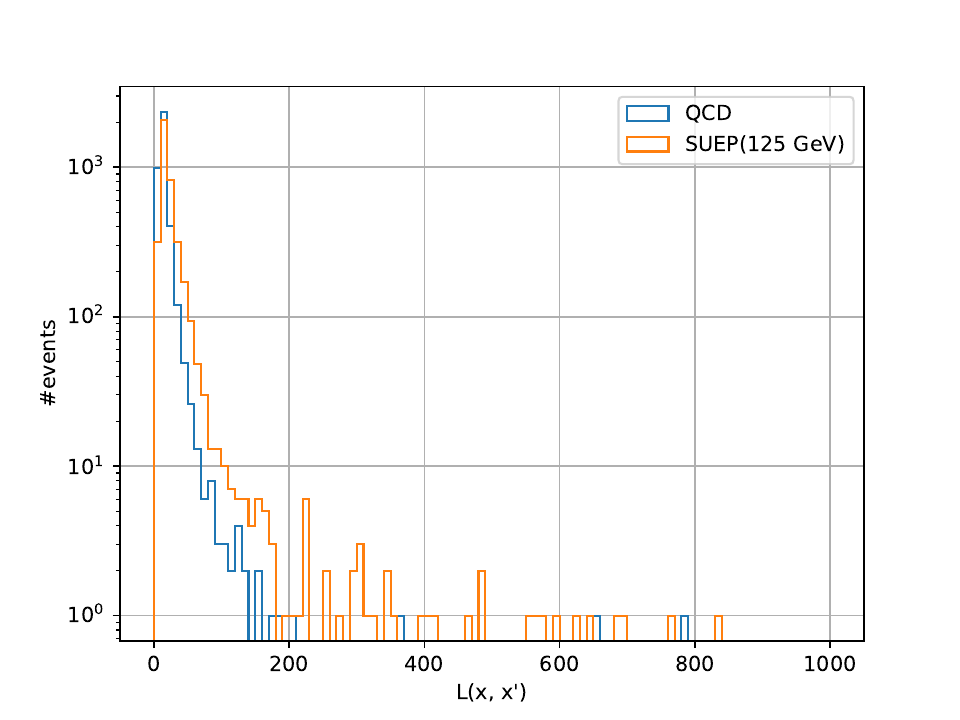}
\includegraphics[width=.4\textwidth]{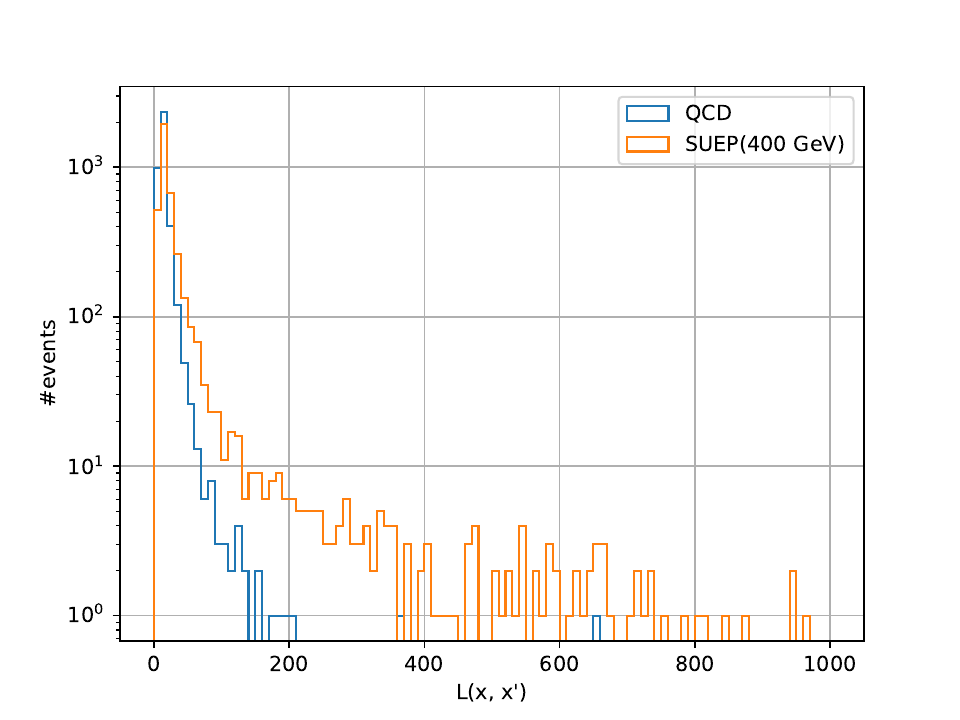}
\includegraphics[width=.4\textwidth]{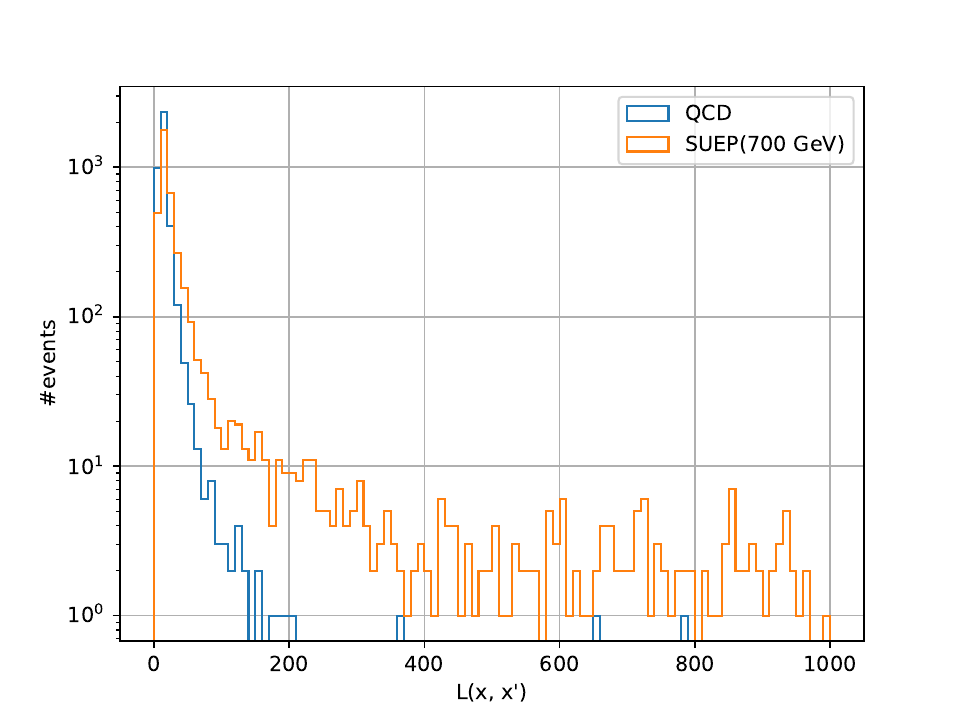}
\includegraphics[width=.4\textwidth]{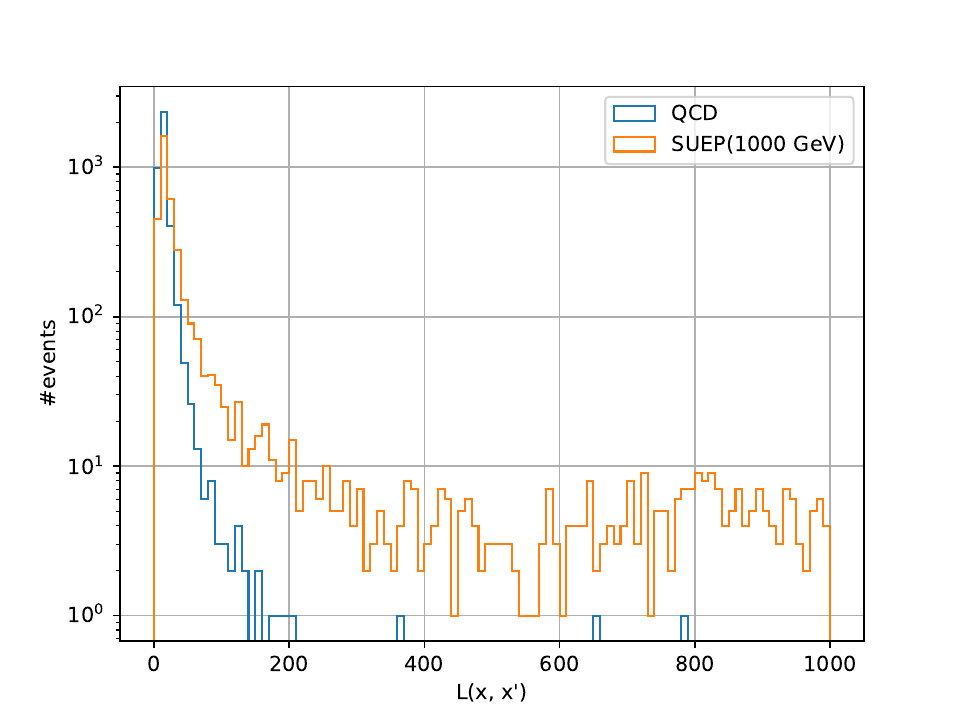}
\caption{Comparisons of ConvAE reconstruction loss for QCD and SUEP(125~\GeV) (top left), SUEP(400~\GeV) (top right), SUEP(700~\GeV) (bottom left), and SUEP(1000~\GeV) (bottom right).}
\label{fig:AElossSUEPvsQCD}
\end{figure}

\begin{figure}[h]
\centering
\includegraphics[width=.4\textwidth]{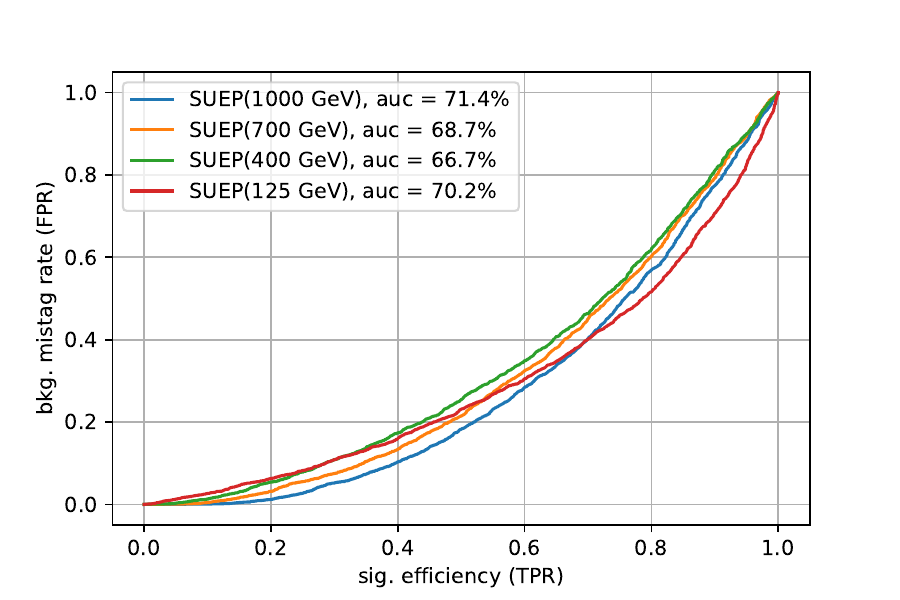}
\includegraphics[width=.4\textwidth]{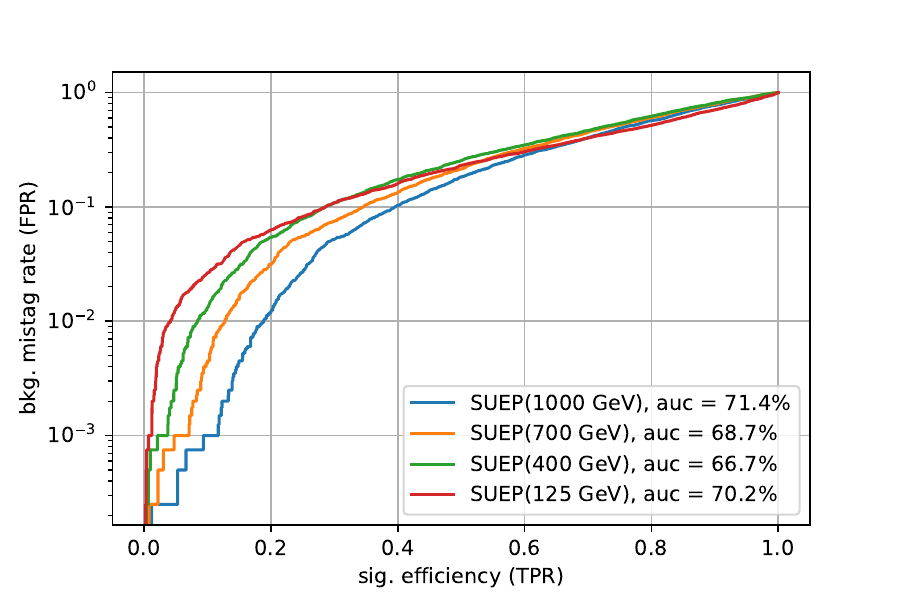}
\caption{ROC curves for ConvAE reconstruction loss as an AD proxy. The right plot is the same as the left plot but with a logarithmic y-scale.}
\label{fig:AElossAsAnomalyDetection}
\end{figure}

The comparisons of ConvAE reconstruction loss for QCD background and SUEP signal with the considered mediator mass scenarios are given in Figure~\ref{fig:AElossSUEPvsQCD}. The corresponding Receiver Operating Characteristic (ROC) curves are given in Figure~\ref{fig:AElossAsAnomalyDetection}. The ConvAE reconstruction loss provides decent discrimination power. The areas under ROC curves (AUCs) are $\sim$70\%, and they barely depend on the mediator mass. For a target of 40\% signal-selection efficiency, the QCD event mistagging rate ranges from 10.3\% for SUEP(1000~\GeV) to 16\% for SUEP(125~\GeV). 

\begin{figure}[h]
\centering
\includegraphics[width=.4\textwidth]{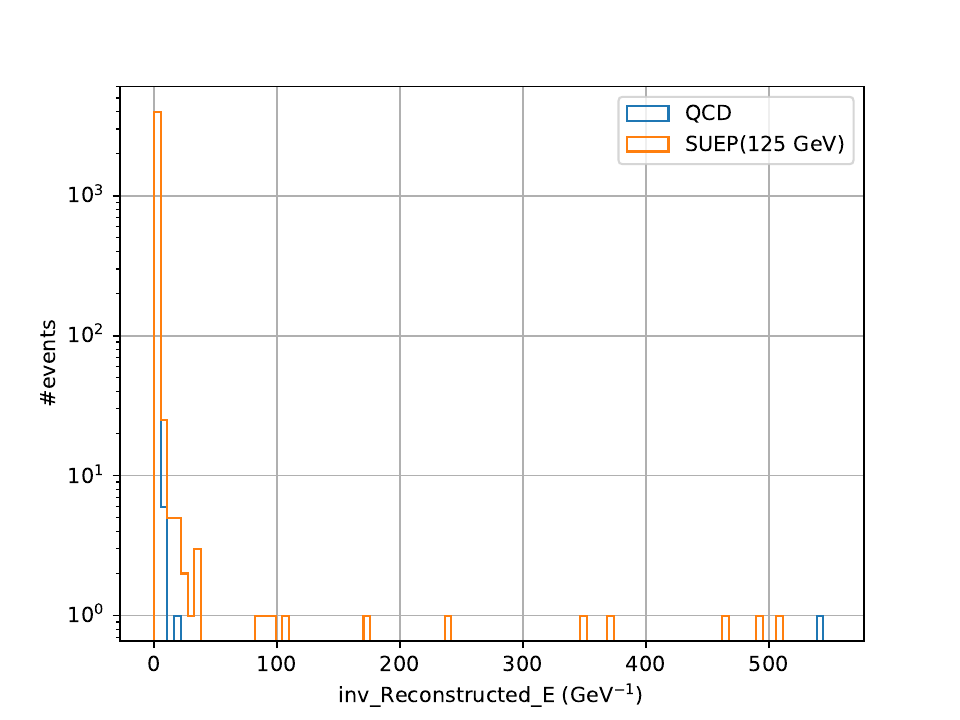}
\includegraphics[width=.4\textwidth]{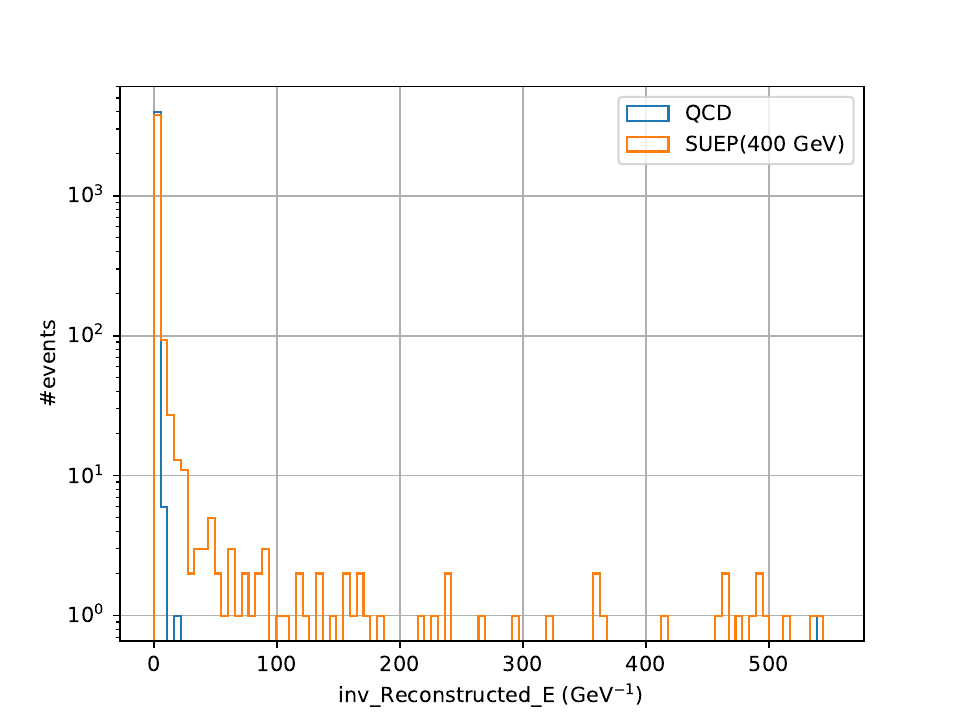}
\includegraphics[width=.4\textwidth]{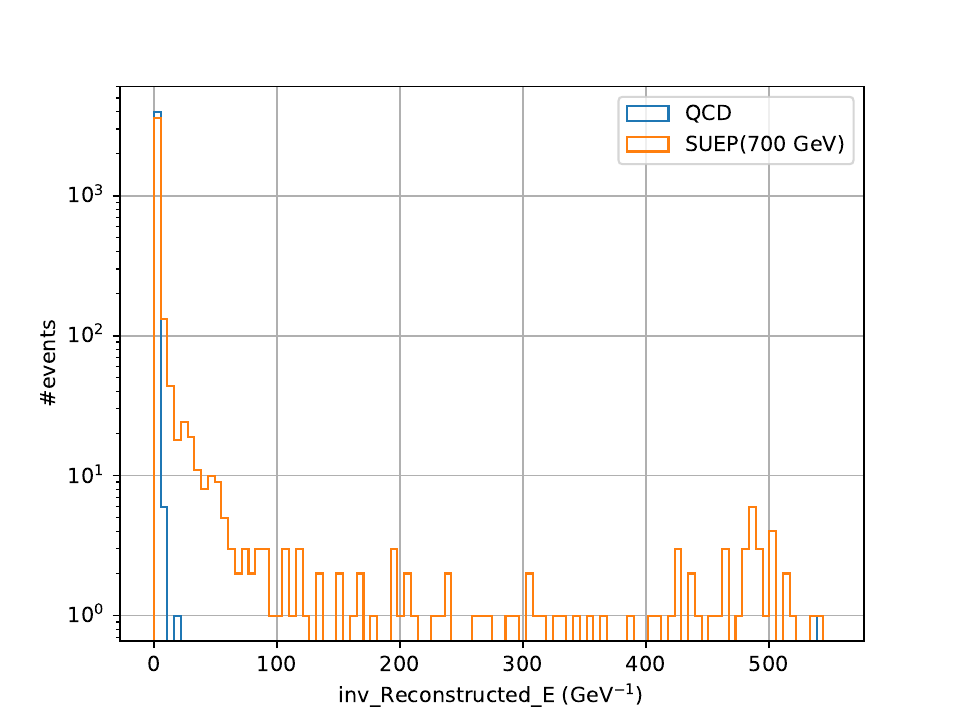}
\includegraphics[width=.4\textwidth]{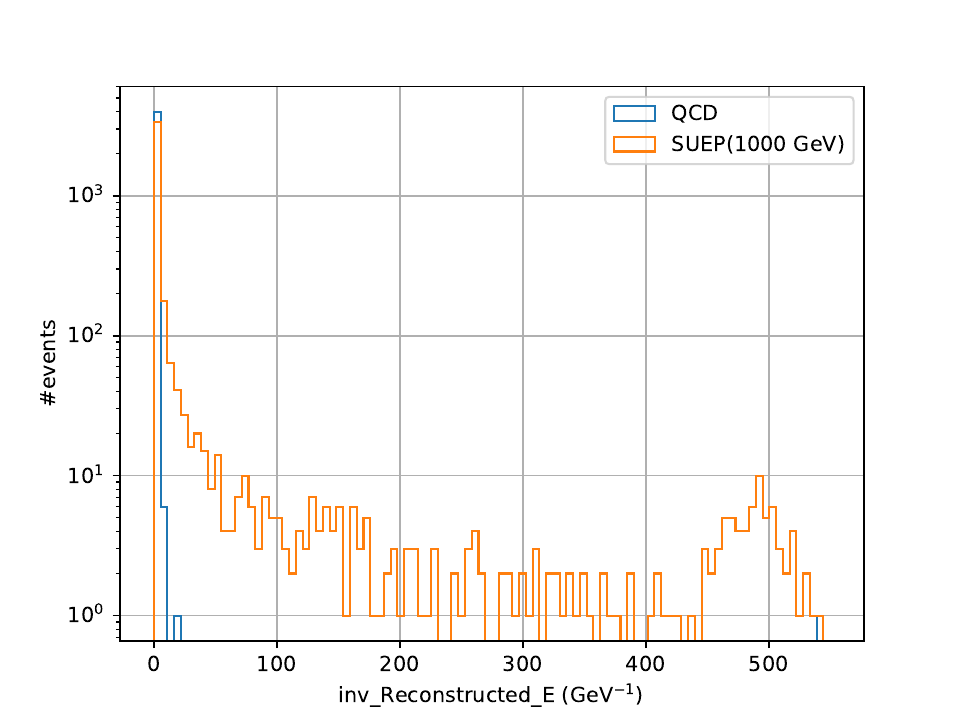}
\caption{Comparisons of the inverse of reconstructed \et for QCD and SUEP(125~\GeV) (top left), SUEP(400~\GeV) (top right), SUEP(700~\GeV) (bottom left), and SUEP(1000~\GeV) (bottom right).}
\label{fig:InvRecoESUEPvsQCD}
\end{figure}

\begin{figure}[h]
\centering
\includegraphics[width=.4\textwidth]{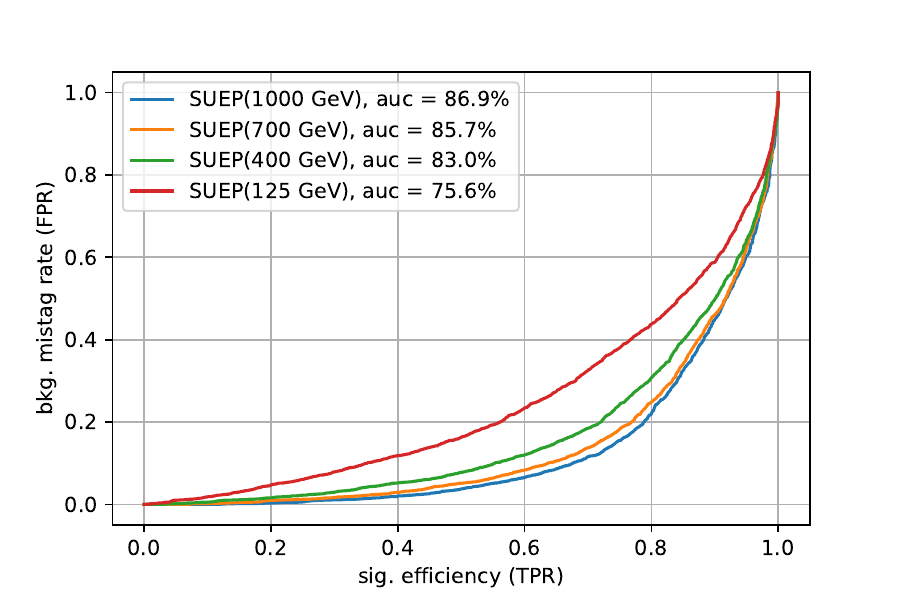}
\includegraphics[width=.4\textwidth]{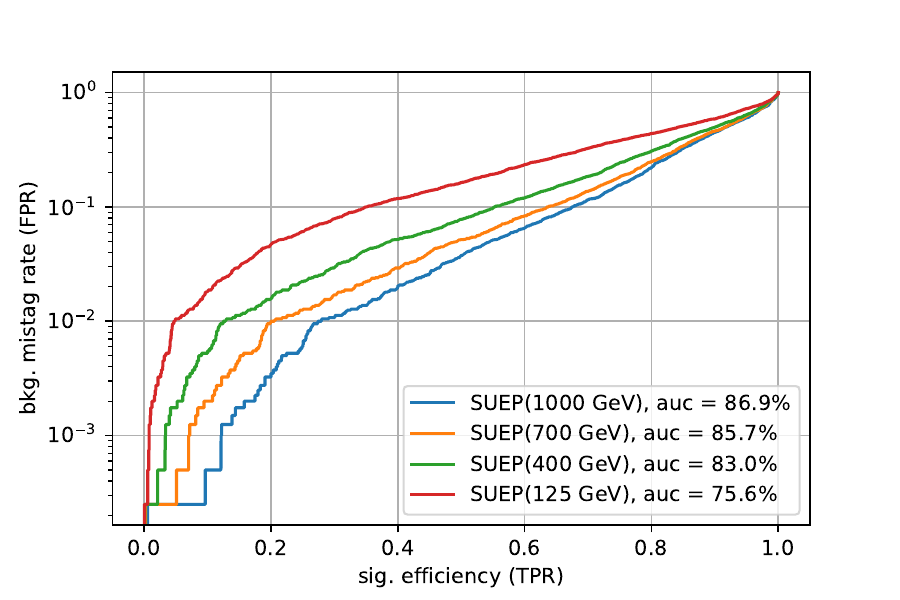}
\caption{ROC curves for the inverse of reconstructed \et as an AD proxy. The right plot is the same as the left plot but with a logarithmic y-scale.}
\label{fig:InvRecoEAsAnomalyDetection}
\end{figure}

We investigated other ConvAE reconstruction parameters to see if they could be used to further enhance the performance. We observed a parameter, the inverse of reconstructed \et, which provides much better discrimination power. The comparisons of the inverse of the reconstructed \et for QCD background and SUEP signal with the considered mediator mass scenarios are given in Figure~\ref{fig:InvRecoESUEPvsQCD}, and the corresponding ROC curves are given in Figure~\ref{fig:InvRecoEAsAnomalyDetection}. The AUCs range from 75.6\% for SUEP(125~\GeV) to 86.9\% for SUEP(1000~\GeV). For a target of 40\% signal-selection efficiency, the QCD mistagging rate ranges from 2\% for SUEP(1000~\GeV) to 11.9\% for SUEP(125~\GeV).

The model inference time has been quantified using a CPU device, an {\texttt{Intel\textregistered~Core$^{\text{TM}}$~i5-9600KF}} processor~\cite{i5processor}, and found to be $\sim$$\SI{20}{\milli\second}$, which is very well within the processing-time limits of the HLT system ($\mathcal{O}(10^2)~\SI{}{\milli\second}$).

\section{Conclusions}
\label{Conclusions}
A deep convolutional neural autoencoder network, trained using background events by taking their signatures in different sub-detectors of an experiment at the Large Hadron Collider as $n$-channel image data, is highly effective as an Anomaly Detection (AD) technique. The training of the autoencoder using unlabeled background events makes this technique highly model-agnostic. We have developed an autoencoder for AD with the Compact Muon Solenoid experiment, which can easily be adapted for other Large Hadron Collider experiments. We have exploited the raw reconstructed signature in the High-Level Trigger system, with the clear benefit of not relying on online particle reconstruction, which could fail to reconstruct new physics signatures.

The autoencoder has been trained using QCD events by taking transverse energy deposits in the inner tracker, electromagnetic calorimeter, and hadron calorimeter sub-detectors as 3-channel image data. The biggest challenge was the sparsity of the data: only $\sim$0.5\% of the total $\sim$$\SI{300}{k}$ pixels per image had non-zero values. The model had a high tendency to learn zeros rather than non-zero values. We explored several loss functions, and ultimately, a non-standard loss function served the purpose---the inverse of the so-called Dice Loss, which contains the pixel-wise multiplication of input and output that forces the output images to have the same non-zero pixels as the input.

The trained model has been tested for Soft Unclustered Energy Patterns (or SUEPs) detection---a new physics signal with an experimental signature of high-multiplicity spherically-symmetric Standard Model particles, anomalous to QCD jets. To check the robustness of the autoencoder, we considered several signal scenarios depending on the new physics model. It has been observed that the model can detect 40\% of the SUEP events at the expense of a QCD event mistagging rate as low as 2\%. The model inference time using the {\texttt{Intel\textregistered~Core$^{\text{TM}}$~i5-9600KF}} processor is measured to be $\sim$$\SI{20}{\milli\second}$, which perfectly satisfies the HLT system's latency requirements.

\section*{Data Availability}

The data used in the creation of this manuscript will be made available upon request.

\section*{Acknowledgements}
S. S. Chhibra, N. Chernyavskaya and B. Maier are supported by CERN, Switzerland under the Senior Research Fellowship Programme. S. S. Chhibra is partly supported by QMUL, UK. B. Maier is partly supported by KIT, Germany. M. Pierini is supported by the European Research Council (ERC) under the European Union's Horizon 2020 research and innovation program (Grant Agreement No. 772369). S. Hasan is supported by SNS di Pisa, Italy and partly supported by ETH Zurich, Switzerland.

\bibliographystyle{unsrt}
\bibliography{References.bib}
\end{document}